\documentclass[twocolumn,amsmath,amssymb]{revtex4}

\begin{document}

\title{{\bf Higher dimensional gravity invariant under the AdS group}}

\author{P. Salgado}
 \email{pasalgad@udec.cl}
\author{F. Izaurieta}
 \email{fizaurie@udec.cl}
\author{E. Rodr\'{\i}guez}
 \email{edurodriguez@udec.cl}
\affiliation{
Departamento de F\'\i sica, Universidad de Concepci\'on,\\
Casilla 160-C, Concepci\'{o}n, Chile.
}%

\begin{abstract}
A higher dimensional gravity invariant both under local Lorentz
rotations and under local Anti de Sitter boosts is constructed. It
is shown that such a construction is possible both when odd
dimensions and when even dimensions are considered. It is also
proved that such actions have the same coefficients as those
obtained in ref. \cite{tron}.

\smallskip\

PACS number(s): 04.50. +h
\end{abstract}

\maketitle

The most general lagrangian for gravity in $d$ dimensions built up
on the same principles as General Relativity (general covariance,
second order equations for the metric, and no explicit torsion) is
a polynomial of degree $\left[ d/2\right] $ in the curvature known
as the Lanczos-Lovelock lagrangian. Lanczos-Lovelock $\left(
\text{LL}\right) $ theories have the same fields, symmetries, and
local degrees of freedom as ordinary gravity. The action can be
written in terms of the Riemann curvature $R^{ab}=d\omega
^{ab}+\omega _{\;c}^a\omega ^{cb}$ and of the vielbein $e^a$
\cite{lanc,lovel,zumino,teit} as
\begin{equation}
S=\int \sum_{p=0}^{\left[ d/2\right] }\alpha _pL^{(p)}  \label{uno}
\end{equation}
where $\alpha _p$ are arbitrary constants and $L^{(p)}$ is given by
\begin{equation}
L^{(p)}=\varepsilon _{a_1a_2\cdot \cdot \cdot \cdot \cdot \cdot
a_d}R^{a_1a_2}\cdot \cdot \cdot \cdot R^{a_{2p-1}a_{2p}}e^{a_{2p+1}}\cdot
\cdot \cdot \cdot e^{a_d}.  \label{dos}
\end{equation}

One important drawback of the Lanczos-Lovelock action is the appearance of a
number of dimensionful constants which are not determined from first
principles. In ref. \cite{tron}\ it is shown that requiring that the
equations of motion uniquely determine the dynamics for as many components
of the independent fields as possible fixes the $\alpha _p$\ coefficients
(for even as well as for odd dimensions) in terms of the gravitational and
cosmological constants.

For $d=2n$ the coefficients are
\begin{equation}
\alpha _p=\alpha _0(2\gamma )^p%
{n \choose p}
,  \label{tres}
\end{equation}
and the action takes a Born-Infeld-like form. With these coefficients, the
LL action is invariant only under local Lorentz rotations.

For $d=2n-1$, the coefficients become
\begin{equation}
\alpha _p=\alpha _0\frac{(2n-1)(2\gamma )^p}{(2n-2p-1)}%
{n-1 \choose p}
,  \label{cuatro}
\end{equation}
where
\begin{equation}
\alpha _0=\frac \kappa {dl^{d-1}},\qquad \gamma =-\text{sgn}(\Lambda )\frac{%
l^2}2,
\end{equation}
and, for any dimension $d$, $l$ is a length parameter related to the
cosmological constant by $\Lambda =\pm (d-1)(d-2)/2l^2.$ With these
coefficients (\ref{cuatro}), the vielbein and the spin connection may be
accommodated into a connection for the AdS group, allowing for the
lagrangian to become the Chern-Simons form in $d=2n+1$ dimensions, whose
exterior derivative is the Euler topological invariant in $d=2n$ dimensions,
\begin{equation}
dL_{CS}^{(2n-1)}=\frac{\kappa l}{2n}\varepsilon _{A_1\cdot \cdot \cdot \cdot
\cdot A_{2n}}{\bf R}^{A_1A_2}\cdot \cdot \cdot {\bf R}^{A_{2n-1}A_{2n}}
\end{equation}
with{\bf \ }
\begin{equation}
{\bf R}^{AB}=\left(
\begin{array}{cc}
R^{ab}+\frac 1{l^2}e^ae^b & \frac 1lT^a \\
-\frac 1lT^b & 0
\end{array}
\right) .
\end{equation}
Thus, the odd-dimensional theory is a gauge theory for the AdS group, and
the independent fields are all components of a connection for this group.

The Chern-Simons construction fails in even-dimensions for the simple reason
that there has not been found a characteristic class constructed with
products of curvature in odd-dimensions. This could be a reason why the
construction of a higher dimensional gravity in even dimensions, invariant
under the AdS group, has remained as an interesting open problem.

It is the purpose of this paper to show that the Stelle-West formalism \cite
{stelle}, which is an application of the theory of nonlinear realizations to
gravity, permits constructing an action for Lanczos-Lovelock gravity theory%
{\bf \ }genuinely invariant under the $AdS$ group. It is shown that such a
construction is possible both when odd dimensions and when even dimensions
are considered. Applications of the theory of non-linear realizations to
gravity have been carried out in different ways in previous research, such
as for example in ref.\cite{Isham} where the vierbein field was considered
as a Goldstone field related to a nonlinear realization of the group $%
GL(4,R) $, or of the affine and conformal groups. In the present work, the
Goldstone fields represent \cite{stelle} a point in an internal AdS space.

The actions invariant under the $AdS$ group are constructed using
the vielbein and spin connection $1$-forms obtained in
ref.\cite{stelle}. It is also proved that such actions have the
same coefficients as those obtained in refs. \cite{tron,salga1}.

The non-linear realizations in de Sitter space can be studied by
the general method developed in ref. \cite{callan,volkov}.
Following these references, we consider a Lie group $G$\ and its
stability subgroup $H.$

The Lie group $G$ has $n$ generators. Let us call $\left\{ {\bf V}_i\right\}
_{i=1}^{n-d}$\ the generators of $H$. We shall assume that the remaining
generators $\left\{ {\bf A}_l\right\} _{l=1}^d$\ are chosen so that they
form a representation of $H.$\ In other words, the commutator $\left[ {\bf V}%
_i,{\bf A}_l\right] $\ should be a linear combination of ${\bf A}_l$\ alone.
A group element $g\in G$\ can be uniquely represented in the form
\begin{equation}
g=e^{\xi ^l{\bf A}_l}h  \label{sw1}
\end{equation}
where $h$\ is an element of $H.$\ The $\xi ^l$\ parametrize the coset space $%
G/H.${\bf \ }We do not specify here the parametrization of $h$.{\bf \ }One
can define the effect of a group element $g_0$\ on the coset space by
\begin{equation}
g_0g=g_0(e^{\xi ^l{\bf A}_l}h)=e^{\xi ^{\prime l}{\bf A}_l}h^{\prime }
\label{sw2}
\end{equation}
or
\begin{equation}
g_0e^{\xi ^l{\bf A}_l}=e^{\xi ^{\prime l}{\bf A}_l}h_1  \label{sw3}
\end{equation}
where
\begin{equation}
h_1=h^{\prime }h^{-1}  \label{sw4}
\end{equation}
\begin{equation}
\xi ^{\prime }=\xi ^{\prime }(g_0,\xi )
\end{equation}
\begin{equation}
h_1=h_1(g_0,\xi ).
\end{equation}

If $g_{0}-1$\ is infinitesimal, (\ref{sw3}) implies
\begin{equation}
e^{-\xi ^{l}{\bf A}_{l}}\left( g_{0}-1\right) e^{\xi ^{l}{\bf A}%
_{l}}-e^{-\xi ^{l}{\bf A}_{l}}\delta e^{\xi ^{l}{\bf A}_{l}}=h_{1}-1.
\label{sw5}
\end{equation}
The right-hand side of (\ref{sw5}) is a generator of $H.$

Let us first consider the case in which $g_{0}=h_{0}\in H.$\ Then (\ref{sw3}%
) gives
\begin{equation}
e^{\xi ^{\prime l}{\bf A}_{l}}=h_{0}e^{\xi ^{l}{\bf A}_{l}}h_{0}^{-1}.
\label{sw6}
\end{equation}
Since the ${\bf A}_{l}$\ form a representation of $H,$\ this implies
\begin{equation}
h_{1}=h_{0};\qquad h^{\prime }=h_{0}h.  \label{sw7}
\end{equation}

The transformation from $\xi $\ to $\xi ^{\prime }$\ given by (\ref{sw6}) is
linear. On the other hand, consider now
\begin{equation}
g_{0}=e^{\xi _{0}^{l}{\bf A}_{l}}.  \label{sw8}
\end{equation}
In this case, eq. (\ref{sw3}) becomes
\begin{equation}
e^{\xi _{0}^{l}{\bf A}_{l}}e^{\xi ^{l}{\bf A}_{l}}=e^{\xi ^{\prime l}{\bf A}%
_{l}}h.  \label{sw9}
\end{equation}
This is a non-linear inhomogeneous transformation for $\xi .$\ The
infinitesimal form of (\ref{sw5}) is
\begin{equation}
e^{-\xi ^{l}{\bf A}_{l}}\xi _{0}^{i}{\bf A}_{i}e^{\xi ^{j}{\bf A}%
_{j}}-e^{-\xi ^{l}{\bf A}_{l}}\delta e^{\xi ^{i}{\bf A}_{i}}=h_{1}-1.
\label{sw10}
\end{equation}
The left-hand side of this equation can be evaluated, using the algebra of
the group. Since the results must be a generator of $H$, one must set equal
to zero the coefficient of $A_{l}.$\ In this way one finds an equation from
which $\delta \xi ^{i}$\ can be calculated.

The construction of a Lagrangian invariant under local group transformations
requires the introduction of a set of gauge fields $a=a_{\mu }^{i}A_{i}$d$%
x^{\mu },$\ $\rho =\rho _{\mu }^{i}V_{i}$d$x^{\mu },$\ $p=p_{\mu }^{l}A_{l}$d%
$x^{\mu }$, $v=v_{\mu }^{i}V_{i}$d$x^{\mu },$\ associated with the
generators $V_{i}$\ and $A_{l}$, respectively. Hence $\rho +a$\ is the usual
linear connection for the gauge group $G,$\ and therefore its transformation
law under $g\in G$\ is
\begin{equation}
g:(\rho +a)\rightarrow (\rho ^{\prime }+a^{\prime })=\left[ g(\rho +a)g^{-1}-%
\frac{1}{f}(dg)g^{-1}\right]  \label{sw12}
\end{equation}
where $f$\ is a constant which, as it turns out, gives the strength of the
universal coupling of the gauge fields to all other fields.

We now consider the Lie algebra valued-differential 1-forms $p$ and $v$
defined by \cite{callan}
\begin{equation}
e^{-\xi ^{l}{\bf A}_{l}}\left[ d+f(\rho +a)\right] e^{\xi ^{l}{\bf A}%
_{l}}=p+v.  \label{sw13}
\end{equation}
The transformation laws for the forms $p(\xi ,d\xi )$\ and $v(\xi ,d\xi )$\
are easily obtained. In fact, using (\ref{sw8}),(\ref{sw9}) one finds \cite
{zumino}
\begin{equation}
p^{\prime }=h_{1}p(h_{1})^{-1}  \label{sw14}
\end{equation}
\begin{equation}
v^{\prime }=h_{1}v(h_{1})^{-1}-\left( dh_{1}\right) h_{1}^{-1}.  \label{sw15}
\end{equation}

Eq. (\ref{sw14}) shows that the differential forms $p(\xi ,d\xi )$\ and $%
v(\xi ,d\xi )$ are transformed linearly by a group element of the form (\ref
{sw8}); the former as a tensor and the latter as a connection. The
transformation law is the same as that by an element of $H$, except that now
this group element $h_1(\xi _0,\xi )$\ is a function of the variable $\xi $.
Therefore, any H-invariant expression, in any dimensions, written with $a$
and $\rho $, will be G-invariant if these fields are changed by $p$ and $v$,
respectively.

We have specified the fields $p$\ and $v$\ as well as their transformation
properties, and now we make use of them to define the covariant derivative
with respect to the group $G$:
\begin{equation}
D=d+v.  \label{sw16}
\end{equation}
The corresponding components of the two-form curvature are
\begin{equation}
T=Dp  \label{sw17}
\end{equation}
\begin{equation}
R=dv+vv.  \label{sw18}
\end{equation}

When $G$ is the AdS lie algebra
\begin{equation}
\left[ P_a,P_b\right] =-im^2J_{ab}
\end{equation}
\begin{equation}
\left[ J_{ab},P_c\right] =i\left( \eta _{ac}P_b-\eta _{bc}P_a\right)
\end{equation}
\begin{equation}
\left[ J_{ab},J_{cd}\right] =i\left( \eta _{ac}J_{bd}-\eta _{bc}J_{ad}+\eta
_{bd}J_{ac}-\eta _{ad}J_{bc}\right)
\end{equation}
having as generators $P_a,J_{ab}$ and the subalgebra $H$ is the Lorentz
algebra $SO(3,1)$ with generators $J_{ab}$, then the equation (\ref{sw13})
becomes
\[
\frac 12iW^{ab}{\bf J}_{ab}-iV^a{\bf P}_a
\]
\begin{equation}
=e^{i\xi ^a{\bf P}_a}\left[ d+\frac 12i\omega ^{ab}{\bf J}_{ab}-ie^a{\bf P}%
_a\right] e^{-i\xi ^b{\bf P}_b}.  \label{sw39}
\end{equation}

Using the AdS algebra, we arrive at explicit expressions for $V^a$ and $%
W^{ab}$ which are the basis for the Stelle-West formalism:

\
\[
V^a=e^a+\left( \cosh z-1\right) \left( \delta _b^a-\frac{\xi _b\xi ^a}{\xi ^2%
}\right) e^b
\]
\begin{equation}
+\frac{senhz}zD\xi ^a-\left( \frac{senhz}z-1\right) \left( \frac{\xi ^cd\xi
_c}{\xi ^2}\xi ^a\right) .  \label{ocho}
\end{equation}

\[
W^{ab}=\omega ^{ab}-\frac 1{l^2}\frac{senhz}z\left( \xi ^ae^b-\xi
^be^a\right)
\]
\begin{equation}
-\frac 1{l^2}\left[ \xi ^aD\xi ^b-\xi ^bD\xi ^a\right] \left( \frac{\cosh z-1%
}{z^2}\right) .  \label{nueve}
\end{equation}
with $z=\frac 1l\left( \xi ^a\xi _a\right) ^{1/2}=\frac 1l\xi .$ Only in the
so called ''physical'' gauge \cite{grigna}, where $\xi ^a=0$, we have $%
V^a=e^a$\ and $W^{ab}=\omega ^{ab}.$ \ In this gauge the resultant
theory is invariant only under the Lorentz group. There is
however, an exceptional case which occurs when the odd-dimensional
case is considered and when the coefficients are appropriate
choices. We shall comment on this below.

On the other hand, under an infinitesimal AdS boost,
\begin{equation}
\delta \xi ^a=\rho ^a+\left( \frac{z\cosh z}{senhz}-1\right) \left( \rho
^a-\rho ^b\frac{\xi _b\xi ^a}{\xi ^2}\right)  \label{seis}
\end{equation}

\begin{equation}
\delta V^a=\overline{\kappa }_b^aV^b  \label{on1}
\end{equation}
\
\begin{equation}
\delta W^{ab}=-{\cal D}\overline{\kappa }^{ab}  \label{on2}
\end{equation}
where
\begin{equation}
\overline{\kappa }^{ab}=-\frac 1{l^2}\frac{\cosh z-1}{z\sinh z}\left( \xi
^a\rho ^b-\xi ^b\rho ^b\right) .
\end{equation}
Then, under the AdS group, $V^a$ and $W^{ab}$ behave as a Lorentz vector and
connection respectively, but with a non-linear parameter.

Now we proceed to apply the above formalism to build a gauge theory of
gravity in any number of dimensions.

The vielbein and the spin connection can be rewritten in the form
\begin{equation}
V^a=O_b^ae^b+D_z\xi ^a,  \label{vier1}
\end{equation}
\[
W^{ab}=\left[ \delta _c^a\delta _g^b-\frac 2{l^2}(\frac{\cosh z-1}{z^2}%
)\delta _{fg}^{ab}\xi ^f\xi _c\right] \omega ^{cg}
\]
\begin{equation}
-\frac 1{l^2}\delta _{fg}^{ab}\left[ (\frac{\cosh z-1}{z^2})\xi ^fD\xi ^g+%
\frac{senhz}z\xi ^fe^g\right] ,  \label{conec}
\end{equation}
where
\begin{equation}
O_b^a\equiv \cosh z\delta _b^a-\left( \cosh z-1\right) \frac{\xi ^a\xi _b}{%
\xi ^2},  \label{vier2}
\end{equation}
\begin{equation}
D_z\equiv \frac{senhz}zD\xi ^a-\left( \frac{senhz}z-1\right) \frac{dz}z.
\label{vier3}
\end{equation}

It is interesting to note that the inverse of the operator $O_b^a$ is given
by
\begin{equation}
\left( O_b^a\right) ^{-1}=\frac 1{\cosh z}\delta _b^a-\left( \frac 1{\cosh z}%
-1\right) \frac{\xi ^a\xi _b}{\xi ^2}.  \label{vier4}
\end{equation}

Under the transformations $e^a\rightarrow e^a+\delta e^a,$ $\omega
^{ab}\rightarrow \omega ^{ab}+\delta \omega ^{ab},$ the vielbein and the
connection change as
\begin{equation}
\delta _eV^a=O_b^a\delta e^b,  \label{vier5}
\end{equation}
\begin{equation}
\delta _\omega V^a=\frac{senhz}z\xi ^b\delta \omega _b^a,  \label{vier6}
\end{equation}
\begin{equation}
\delta _eW^{ab}=-\frac 1{l^2}\frac{senhz}z\delta _{fg}^{ab}\xi ^f\delta e^g,
\label{vier7}
\end{equation}
\begin{equation}
\delta _\omega W^{ab}=\delta \omega ^{ab}-\left( \cosh z-1\right) \delta
_{fg}^{ab}\frac{\xi ^f\xi _c}{\xi ^2}\delta \omega ^{cg}.  \label{vier8}
\end{equation}

The action (\ref{uno}) can now be written in the form

\[
S=\int \sum_{p=0}^k\alpha _p\varepsilon _{a_1a_2\cdot \cdot \cdot \cdot
\cdot \cdot a_d}R^{a_1a_2}\cdot \cdot \cdot \cdot
\]
\begin{equation}
\cdot \cdot \cdot \cdot \cdot R^{a_{2p-1}a_{2p}}V^{a_{2p+1}}\cdot \cdot
\cdot \cdot V^{a_d},  \label{veinticinco}
\end{equation}
where now
\begin{equation}
R^{ab}=dW^{ab}+W_c^aW^{cb}.  \label{veintiseis}
\end{equation}
The space-time torsion ${\cal T}^{\text{ }a}$ is given by
\begin{equation}
{\cal T}^{\text{ }a}={\cal D}V^a
\end{equation}
where ${\cal D}$ is the covariant derivative in the connection $W^{ab}.$

This action (\ref{veinticinco}) is invariant under general coordinate
transformations and under $AdS$ transformations (\ref{on1}), (\ref{on2}).
The interesting result is that the action (\ref{veinticinco}) is invariant
under $AdS$ transformations, not only for the odd-dimensional case $d=2n-1$,
but also for the even-dimensional case $d=2n.$

Now we consider the variations of the action with respect to $\xi ^a$, $e^a$%
, $\omega ^{ab}.$ The variations of the action (\ref{veinticinco}) with
respect to $e^a,$ $\omega ^{ab}$ lead to the following equations:
\[
\sum_{p=0}^{\left[ d-1/2\right] }\frac 2{l^2}\frac{senhz}zp(d-2p)\alpha
_p\varepsilon _{a_1\cdot \cdot \cdot \cdot \cdot \cdot a_d}\xi ^{a_1}{\cal T}%
\text{ }^{a_2}
\]
\[
\times R^{a_3a_4}\cdot \cdot \cdot \cdot \cdot
R^{a_{2p-1}a_{2p}}V^{a_{2p+1}}\cdot \cdot \cdot \cdot V^{a_{d-1}}
\]
\[
+\sum_{p=0}^{\left[ d-1/2\right] }(d-2p)\alpha _p\varepsilon _{a_1\cdot
\cdot \cdot \cdot \cdot \cdot a_{d-1}f}R^{a_1a_2}\cdot \cdot \cdot
\]
\begin{equation}
\cdot \cdot \cdot R^{a_{2p-1}a_{2p}}V^{a_{2p+1}}\cdot \cdot \cdot \cdot
V^{a_{d-1}}O_{a_d}^f=0.  \label{LL1}
\end{equation}

\[
\sum_{p=0}^{\left[ d-1/2\right] }p(d-2p)\alpha _p\varepsilon _{a_1\cdot
\cdot \cdot \cdot \cdot \cdot a_d}V^{a_1}{\cal T}\text{ }^{a_2}
\]
\[
\times R^{a_3a_4}\cdot \cdot \cdot \cdot R^{a_{2p-1}a_{2p}}V^{a_{2p+1}}\cdot
\cdot \cdot \cdot V^{a_{d-2}}
\]
\[
+\sum_{p=0}^{\left[ d-1/2\right] }\frac{2(\cosh z-1)}{l^2z^2}p(d-2p)\alpha
_p\varepsilon _{a_1\cdot \cdot \cdot \cdot \cdot a_{d-2}fa_d}\xi ^{a_1}{\cal %
T}\text{ }^{a_2}
\]
\[
\times R^{a_3a_4}\cdot \cdot \cdot \cdot \cdot \cdot
R^{a_{2p-1}a_{2p}}V^{a_{2p+1}}\cdot \cdot \cdot V^{a_{d-1}}V^f\xi _{a_{d-1}}
\]
\[
+\sum_{p=0}^{\left[ d-1/2\right] }\frac{senhz}z(d-2p)\alpha _p\varepsilon
_{a_1\cdot \cdot \cdot \cdot \cdot a_{d-2}fa_d}R^{a_1a_2}\cdot \cdot \cdot
\]
\begin{equation}
\cdot \cdot \cdot \cdot R^{a_{2p-1}a_{2p}}V^{a_{2p+1}}\cdot \cdot \cdot
V^{a_{d-1}}V^f\xi _{a_{d-1}}=0.  \label{LL2}
\end{equation}

These equations reproduce the equations of motion of Lanczos-Lovelock
gravity theory. In fact, taking the product of Eq. (\ref{LL1}) with ($%
O_b^a)^{-1},$ and of Eq. (\ref{LL2}) with $\xi ^{a_{d-1}},$ we obtain
\[
\sum_{p=0}^{\left[ d-1/2\right] }\frac 2{l^2z}\tanh z\text{ }p(d-2p)\alpha
_p\varepsilon _{a_1\cdot \cdot \cdot \cdot \cdot \cdot a_{d-1}a_d}\xi ^{a_1}%
{\cal T}\text{ }^{a_2}
\]
\[
\times R^{a_3a_4}\cdot \cdot \cdot \cdot R^{a_{2p-1}a_{2p}}V^{a_{2p+1}}\cdot
\cdot \cdot \cdot V^{a_{d-1}}
\]
\[
+\sum_{p=0}^{\left[ d-1/2\right] }(d-2p)\alpha _p\varepsilon _{a_1\cdot
\cdot \cdot \cdot \cdot \cdot a_{d-1}a_d}R^{a_1a_2}\cdot \cdot \cdot
\]
\begin{equation}
\cdot \cdot \cdot \cdot R^{a_{2p-1}a_{2p}}V^{a_{2p+1}}\cdot \cdot \cdot
\cdot V^{a_{d-1}}=0.  \label{LL3}
\end{equation}

\[
\sum_{p=0}^{\left[ d-1/2\right] }\frac{(2\cosh z-3)}{l^2zsenhz}\text{ }%
p(d-2p)\alpha _p\varepsilon _{a_1\cdot \cdot \cdot \cdot \cdot \cdot a_d}\xi
^{a_1}{\cal T}\text{ }^{a_2}
\]
\[
\times R^{a_3a_4}\cdot \cdot \cdot R^{a_{2p-1}a_{2p}}V^{a_{2p+1}}\cdot \cdot
\cdot \cdot V^{a_{d-1}}
\]
\[
-\sum_{p=0}^{\left[ d-1/2\right] }(d-2p)\alpha _p\varepsilon _{a_1\cdot
\cdot \cdot \cdot \cdot \cdot \cdot a_d}R^{a_1a_2}\cdot \cdot \cdot \cdot
\]
\begin{equation}
\cdot \cdot \cdot \cdot R^{a_{2p-1}a_{2p}}V^{a_{2p+1}}\cdot \cdot \cdot
V^{a_{d-1}}=0.  \label{LL4}
\end{equation}

Taking the addition of Eq. (\ref{LL3}) to Eq. (\ref{LL4}), we have
\[
\sum_{p=0}^{\left[ d-1/2\right] }\text{ }p(d-2p)\alpha _p\varepsilon
_{a_1\cdot \cdot \cdot \cdot \cdot \cdot a_d}\xi ^{a_1}{\cal T}\text{ }%
^{a_2}
\]
\begin{equation}
\times R^{a_3a_4}\cdot \cdot \cdot R^{a_{2p-1}a_{2p}}V^{a_{2p+1}}\cdot \cdot
\cdot \cdot V^{a_{d-1}}=0,  \label{LL5}
\end{equation}
and therefore we can write
\[
\sum_{p=0}^{\left[ d-1/2\right] }(d-2p)\alpha _p\varepsilon _{a_1\cdot \cdot
\cdot \cdot \cdot \cdot \cdot a_d}R^{a_1a_2}\cdot \cdot \cdot
\]
\begin{equation}
\cdot \cdot \cdot \cdot R^{a_{2p-1}a_{2p}}V^{a_{2p+1}}\cdot \cdot \cdot
V^{a_{d-1}}=0.  \label{LL6}
\end{equation}

From (\ref{LL6}), (\ref{LL3}) and (\ref{LL2}) we obtain:
\[
\sum_{p=0}^{\left[ d-1/2\right] }p(d-2p)\alpha _p\varepsilon _{aba_3\cdot
\cdot \cdot \cdot \cdot \cdot a_d}R^{a_3a_4}\cdot \cdot \cdot
\]
\begin{equation}
\cdot \cdot \cdot R^{a_{2p-1}a_{2p}}{\cal T}\text{ }^{a_{2p+1}}V^{a_{2p+2}}%
\cdot \cdot \cdot \cdot V^{a_d}=0.  \label{LL7}
\end{equation}

These equations have the same form as those obtained for Lanczos-Lovelock
theory with the usual fields $e^a$ and $\omega ^{ab}$ replaced by $V^a$ and $%
W^{ab}$. The field equation corresponding to the variation of the action
with respect to $\xi ^a,$ is not an independent equation. In fact, taking
the covariant derivative operator ${\cal D}$ of equation (\ref{LL1}), we
obtain the same equation that one obtains by varying the action with respect
to $\xi ^a.$ This is to be expected, since the Goldstone field $\xi ^a$ has
no dynamical degrees of freedom.

Following the same procedure of ref. \cite{tron} one can see that the
equations (\ref{LL6}), (\ref{LL7}), lead, in the even-dimensional case, to
the coefficients given by eq.(\ref{tres}) and, in the odd-dimensional case,
to the coefficients given by eq.(\ref{cuatro}).

Therefore, the use of the $SW$-formalism does not change the coefficients $%
\alpha _p$ of the action already obtained in
refs.\cite{tron,salga1}.
In these refs. can be found a more detailed discussion of the coefficients $%
\alpha _p.$

We have shown in this Letter that the Stelle-West formalism for
non-linear gauge theories allows the construction of an off-shell
AdS-invariant higher dimensional gravity. No boundary terms are
added to the lagrangian when the AdS gauge transformations are
performed. This is accomplished in any dimensions, no matter
whether it be even or odd. It is also proved that such actions
have the same coefficients as those obtained in refs.
\cite{tron,salga1}.

We emphasize that the action (\ref{veinticinco}) is invariant under the AdS
group and that,when one picks the physical gauge $\xi ^a=0,$ the theory
becomes indistinguishable from the usual one, and the AdS symmetry is broken
down to the Lorentz group. The only exception to this rule occurs in odd
dimensions when the coefficients (\ref{cuatro})\ are chosen. In this case,
and for any value of $\xi ^a,$ it is possible to show that the
Euler-Chern-Simons action written with $e^a$ and $\omega ^{ab}$ differs from
that written with $V^a$ and $W^{ab}$ by a boundary term. As a matter of
fact, the defining relation for the non-linear fields $V^a$ and $W^{ab}$
given in (\ref{sw39}), represents a gauge transformation for the linear
connection ${\bf A}=\frac 12i\omega ^{ab}{\bf J}_{ab}-ie^a{\bf P}_a$, which
can be written in the form
\begin{equation}
{\bf A\rightarrow \tilde{A}}=g^{-1}\left( d+{\bf A}\right) g,
\end{equation}
where $g=e^{-i\xi ^aP_a}$ and ${\bf \tilde{A}}=\frac 12iW^{ab}{\bf J}%
_{ab}-iV^a{\bf P}_a$. This means that the linear and non-linear curvatures $%
{\bf F}=d{\bf A}+{\bf A}^2$ and ${\bf \tilde{F}}=d{\bf \tilde{A}}+{\bf
\tilde{A}}^2$ are related by
\begin{equation}
{\bf \tilde{F}}=g^{-1}{\bf F}g.  \label{izau}
\end{equation}
Just as the usual Euler-Chern-Simons lagrangian, the odd-dimensional
non-linear lagrangian, with the special choice of coefficients given in eq. (%
\ref{cuatro}), satisfies
\begin{equation}
dL_{\text{VW}}^{\left( 2n-1\right) }=\left\langle {\bf \tilde{F}}%
^n\right\rangle ,
\end{equation}
where $\left\langle {\bf J}_{a_1a_2}\cdots {\bf J}_{a_{2n-3}a_{2n-2}}{\bf P}%
_{a_{2n-1}}\right\rangle =\frac 1l\varepsilon _{a_1\cdots a_{2n-1}}$. Then,
eq. (\ref{izau}) implies that
\begin{equation}
dL_{\text{VW}}^{\left( 2n-1\right) }=\left\langle {\bf \tilde{F}}%
^n\right\rangle =\left\langle {\bf F}^n\right\rangle =dL_{\text{CS}}^{\left(
2n-1\right) },
\end{equation}
and hence we see that both lagrangians may locally differ only by a total
derivative. The same arguments lead to the conclusion that, in general, any
Chern-Simons lagrangian written with non-linear fields differs from the
usual one by a total derivative.

When the linear lagrangian is invariant only under the stability subgroup,
as it happens in even dimensions, the introduction of the non-linear fields
brings in new terms in the action, which cannot be written as boundary
contributions. We are left with a new action which is invariant under the
full group.

It is perhaps interesting to note that, if one considers $g_{\mu
\nu }=V_\mu ^aV_\nu ^b\eta _{ab}$, one can write the lagrangian of
the action (\ref {veinticinco}) in the form that was written in
ref. \cite{teit}. This means that, if one considers the theory to
be constructed in terms of the space-time metric $g_{\mu \nu }$,
ignoring the underlying formulation, the theory described in our
manuscript is completely equivalent to the theory developed in
refs.\cite{lovel,teit}. No trace of the new structure of the
vierbein existing in the underlying formulation of the theory can
be found at the metric level.

The interest in the study of{\bf \ }the Lanczos-Lovelock gravity
theory invariant under $AdS$ transformations takes root in the
fact that,{\bf \ }in recent years, M theory has become the
preferred description for the underlying structure of string
theory \cite{Witt,Sch}. Some of the expected features of M-theory
are (i) its dynamics should somehow exhibit a superalgebra in
which the anticommutator of two supersymmetry generators coincides
with the $AdS$ superalgebra in eleven dimensions \cite{town}, (ii)
the low-energy regime should be described by an eleven dimensional
supergravity of a new type which should stand on a firm geometric
foundation in order to have an off-shell local supersymmetry
\cite{nish}, (iii) the perturbation expansion for graviton
scattering in M-theory has recently led to a conjecture that the
new supergravity Lagrangian should contain higher powers of
curvature \cite{green}.

\begin{acknowledgments}

This work was supported in part by Direcci\'{o}n de
Investigaci\'{o}n, Universidad de Concepci\'{o}n through Grant
N$_o$. 202.011.031-1.0. (P.S) wishes to thank Hermann Nicolai for
the hospitality in AEI in Golm bei Potsdam where part of this work
was done, and to Deutscher Akademischer Austauschdienst (DAAD) for
financial support. (F.I) and (E.R) wish to thank Ricardo Troncoso
and Jorge Zanelli for their warm hospitality in Valdivia and for
enlightening discussions. The authors are grateful to Universidad
de Concepci\'{o}n for partial support of the 2$^{nd}$ Dichato
Cosmological Meeting, where this work was started.

\end{acknowledgments}

\end{document}